\documentclass[12pt,preprint2]{aastex}
\usepackage{epsfig}
\begin{document}

\title{Candidate Type II Quasars from the Sloan Digital Sky Survey: \\
II. From Radio to X-Rays}

\author{
Nadia L. Zakamska\altaffilmark{1}, 
Michael A. Strauss\altaffilmark{1}, 
Timothy M. Heckman\altaffilmark{2}, \\
\v{Z}eljko Ivezi\'c\altaffilmark{1}, 
Julian H. Krolik\altaffilmark{2}\\  
\altaffiltext{1}{Princeton University Observatory, Princeton, New Jersey 08544}
\altaffiltext{2}{Department of Physics and Astronomy, Johns Hopkins University, 3400 North Charles Street, Baltimore, MD 21218-2686}
}
\footnotesize

\begin{abstract}
Type II quasars are luminous AGNs whose central engines and broad-line regions are obscured by intervening material; such objects only recently have been discovered in appreciable numbers. We study the multiwavelength properties of 291 type II AGN candidates ($0.3<z<0.8$) selected based on their optical emission line properties from the spectroscopic database of the Sloan Digital Sky Survey. This sample includes about 150 objects luminous enough to be classified as type II quasars. We matched the sample to the FIRST (20 cm), IRAS (12$-$100\micron), 2MASS ($JHK_S$) and RASS (0.1$-$2.4 keV) surveys. Roughly 10\% of optically selected type II AGN candidates are radio-loud, comparable to the AGN population as a whole. About 40 objects are detected by IRAS at 60\micron\ and/or 100\micron, and the inferred mid/far-IR luminosities lie in the range $\nu L_{\nu}=10^{45}-3\times 10^{46}$ erg sec$^{-1}$. Average IR-to-[OIII]5007 ratios of objects in our sample are consistent with those of other AGNs. Objects from our sample are ten times less likely to have soft X-ray counterparts in RASS than type I AGNs with the same redshifts and [OIII]5007 luminosities. The few type II AGN candidates from our sample that are detected by RASS have harder X-ray spectra than those of type I AGNs. The multiwavelength properties of the type II AGN candidates from our sample are consistent with their interpretation as powerful obscured AGNs.
\end{abstract}

\keywords{galaxies: active --- galaxies: quasars: general --- infrared: galaxies --- surveys --- X-rays: galaxies}

\pagebreak[4]
\section{Introduction}

Type II quasars are luminous analogs of Seyfert 2 galaxies; they were predicted by unification models of active galactic nuclei (AGNs) as powerful objects whose central regions are shielded from the observer by large amounts of gas and dust \citep{anto93, urry95}. However, type II quasars have turned out to be hard to find. Previous searches have defined a total of a few dozen candidate type II quasars at a variety of wavelengths, and the multiwavelength properties of this heterogeneous compilation do not always show clear trends. 

Narrow-line radio galaxies have been known for decades, and powerful central engines (up to $10^{46}-10^{47}$ erg sec$^{-1}$) are invoked in order to explain their narrow-line luminosities \citep{mcca93}. High sensitivity X-ray observations have provided direct evidence for quasar nuclear luminosities and large circumnuclear obscuration ($N_H\sim 10^{22}-10^{24}$ cm$^{-2}$) in some of these objects (e.g., \citealt{samb99, cari02, derr03}). However, this is not a universal result, and some narrow-line radio galaxies show only small absorbing columns, similar to those of type I AGNs \citep{samb99, basi02, barc03}. To further complicate matters, the radio jets themselves can be powerful X-ray emitters, and this component has to be disentangled from the nuclear emission in order to understand the energetics of these objects (e.g., \citealt{youn02}). Because radio-loud objects constitute only a small fraction of the AGN population and the exact radio-loud to radio-quiet ratio is not well constrained for high luminosities, it is difficult to use narrow-line radio galaxy samples for demographic studies.

A large sample (358 objects) of spectroscopically confirmed AGNs selected from the IRAS Point Source Catalog was presented by \citet{degr92}. This sample is important for studies of obscured AGNs since most of their observed luminosity is emitted at mid-infrared. Due to the IRAS sensitivity limit this sample probes only a small cosmological volume (93\% of all AGNs in this sample have redshifts $z<0.1$) and therefore contains only a few objects with quasar luminosities. 

Interest in type II quasars has increased since the discovery that the X-ray background is largely due to obscured AGNs at redshifts around $z\sim 1$ (\citealt{barg03} and references therein), and several candidates were identified. The connection between X-ray selected obscured AGNs and optically identified narrow emission line AGNs remains unclear, as in some cases sources classified as type I AGNs in the optical appear to be absorbed in the X-ray band (see \citealt{matt02} for review). In addition, most Compton-thick sources ($N_H>10^{23.5}$ cm$^{-2}$) are very faint at $0.1-10$ keV, and therefore are not present in the X-ray selected samples. More than 1/3 of Seyfert 2 galaxies have $N_H>10^{24}$ cm$^{-2}$ \citep{maio98, bass99}, but this fraction is not known for objects with quasar luminosities. 

Type II AGNs are not easily found in optical surveys. By analogy with classical Seyfert 2 galaxies, type II quasars are defined as high-ionization narrow emission line objects. Unlike type I quasars which are recognized by their UV excess, type II quasars do not show distinctive colors. Type II AGNs are also much fainter in optical bands than type I AGNs of the same intrinsic luminosity, and therefore deeper surveys are needed to find them in appreciable numbers. Several candidates at $z\sim 0.3$ were selected from the Palomar Sky Survey \citep{djor01}. In a previous paper (\citealt{z03}, hereafter Paper I) we presented a sample of 291 type II AGNs in the redshift range $0.30<z<0.83$ selected from the spectroscopic database of the Sloan Digital Sky Survey (SDSS, \citealt{york00, stou02}) based on their emission line properties. These objects have narrow emission lines ($FWHM<1000$ km sec$^{-1}$), with high-ionization line ratios that clearly distinguish them from other types of emission line objects. They typically have very weak continua which are often dominated by the light from the host galaxy. Both the broad-line region and the UV continuum are completely obscured at optical wavelengths. We use the [OIII]5007 emission line as a tracer of the nuclear activity, as it is emitted from the extended (and therefore presumably less obscured) narrow-line region. By studying [OIII]5007 emission line luminosities, we found that about 50\% of the objects in our sample are luminous enough to be classified as type II quasars. The sample of type II quasars selected from the SDSS is several times larger than the total number of previously known candidates. 

Searches for type II quasars at different wavelengths do not retrieve the same population of objects, and it is far from clear how different selection techniques relate to each other. Therefore, the contribution from obscured AGNs to the black hole census \citep{yu02}, their luminosity function and their fraction in the total AGN population are unknown. Whether or not the unification models developed for low-luminosity Seyfert galaxies apply to AGNs with quasar luminosities is also unknown since there do not exist enough observational data on type II AGNs with high luminosities. 

This paper is the first step in our program to follow up the optically selected sample of type II AGNs at different wavelengths; we hope to be able to resolve many of the issues discussed above with a large, well-defined sample with known multi-wavelength properties. In Sections \ref{radio}$-$\ref{x} we study the radio, mid-infrared, near-infrared and soft X-ray emission of the objects in our sample, respectively, and discuss our results in Section \ref{conclusions}. Throughout this paper we use an $h=0.7$, $\Omega_m=0.3$, $\Omega_{\Lambda}=0.7$ cosmology and J2000 coordinates to identify the objects. Optical emission lines are identified by their air wavelengths in \AA\ (e.g., [OIII]5007).

\section{Radio observations}
\label{radio}

\subsection{Positional matching to the \\FIRST catalog}

The Faint Images of the Radio Sky at Twenty cm survey (FIRST, \citealt{beck95}) is using the Very Large Array (VLA) to produce a map of the 20cm (1.4 GHz) sky with 1.8\arcsec\ pixels, a typical rms of 0.15 mJy, a resolution of 5\arcsec\ and sub-arcsec positional accuracy. The FIRST catalog \citep{whit97} currently includes about 811,000 sources from 9030 square degrees down to the source detection threshold of about 1 mJy; we used the most recent version of the catalog released in April 2003. The coverage of the FIRST survey was chosen to follow the SDSS footprint, so most objects in our sample have FIRST coverage. 

Fields of 21 objects in our sample were not observed by the FIRST survey. Of the remaining 270 type II AGNs, 136 objects have matches in the FIRST catalog within 3\arcsec. The 3\arcsec\ matching radius includes almost all true matches and is essentially contamination-free \citep{ivez02}. This matching procedure finds radio counterparts with complex morphology as long as there is a core component, but extended radio sources with faint core component would not be recognized. Therefore we also visually examined 5\arcmin$\times$5\arcmin\ FIRST fields centered on the objects from our sample. At the redshifts of the objects in our sample ($0.3<z<0.8$), the search size of 2.5\arcmin\ translates into a physical distance of 0.7-1.2 Mpc from the optical source; this radius is thus large enough to cover most known extragalactic radio jets \citep{kais99} and physically associated radio sources \citep{mcma01, ivez02}. An additional 7 matches were found through this visual inspection.

\begin{figure}
\epsscale{1.0}
\plotone{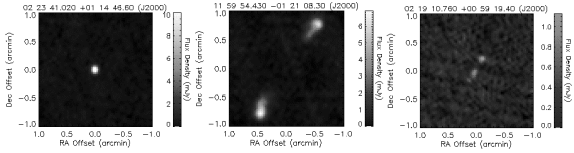}
\figcaption{Examples of high-confidence matches. From left to right: a point source brighter than 1mJy within 3\arcsec\ of the SDSS position; a double-lobed source symmetric around the SDSS position; and an asymmetric source whose core component is centered on the SDSS position. The size of all stamps is 2\arcmin$\times$2\arcmin; they are centered on the SDSS positions (given on top of each stamp); the vertical bar on the right gives the greyscale code for brightness.   \label{confidence-matches}}
\end{figure}

In order to avoid matching with unrelated objects that happen to lie in the 5\arcmin$\times$5\arcmin\ FIRST fields, we include only three kinds of radio morphologies, examples of which are shown in Figure \ref{confidence-matches}. SDSS J022341.02$+$011446.6 has an unresolved radio counterpart (Figure \ref{confidence-matches} left) recognized by the 3\arcsec\ matching algorithm. SDSS J115954.43$-$012108.3 has a symmetric double-lobed radio counterpart (Figure \ref{confidence-matches} middle) centered on the optical position. In this particular object, the core component is not detected and the object was not matched within 3\arcsec. SDSS J021910.76$+$005919.4 has an asymmetric radio counterpart (Figure \ref{confidence-matches} right). Asymmetric objects are included only if they have a core component. Overall, there are 17 objects with radio counterparts which appear extended in FIRST images. Two of them (described above) are shown in Figure \ref{confidence-matches}, and the remaining 15 are shown in Figure \ref{extended}.

In Table 1 we summarize optical and radio information for the 143 objects with FIRST counterparts. Redshifts and [OIII]5007 line luminosities are taken from Paper I. As a measure of total radio flux $F_{\nu}$, for point sources we use the integrated flux $F_{int}$ from the FIRST catalog. For extended sources, we sum up $F_{int}$ for all components of the radio counterpart. The core flux is set to 0 if the core component is not detected (e.g., in SDSS J115954.43$-$012108.3, Figure \ref{confidence-matches} middle) or if it cannot be disentangled from the extended emission at the FIRST resolution (e.g., in SDSS J090933.51$+$425346.5, Figure \ref{extended}, third row, left). Spectral indices are available only for objects with matches in other surveys (see Section \ref{radio_other}). Core fluxes from FIRST were already published for our sample in Paper I. 

For some extended sources, FIRST underestimates the total flux due to overresolution \citep{whit97}. We therefore matched our sample to the NRAO VLA Sky Survey (NVSS; \citealt{cond98}). NVSS is a 1.4 GHz survey covering the entire sky north of declination $-40^o$ with a FWHM resolution of 45\arcsec\ and positional accuracy of better than 7\arcsec. The NVSS catalog\footnote{http://www.cv.nrao.edu/nvss/} contains almost 2 million sources down to about 2.3 mJy sensitivity for point sources. If the total NVSS flux exceeded the total FIRST flux by more than one standard deviation, we listed the NVSS total flux in Table 1. In these cases, a reference to \citet{cond98} was given in the last column of Table 1. Several sources which appear point-like in FIRST images show faint extended emission in NVSS images. Although these sources are not shown in Figures \ref{confidence-matches} and \ref{extended}, they are classified as extended in Table 1.

\subsection{Matching to other radio surveys}
\label{radio_other}

We matched the 21 sources without FIRST coverage to the NVSS \citep{cond98} and we found four matches within 30\arcsec. In addition, the NVSS image of the field of SDSS J121637.25$+$672441.5 (Appendix \ref{SDSS1216}) shows two aligned radio sources which are nearly symmetric around the optical position and with the peak separation of 4.6\arcmin. The five sources with NVSS matches are listed in Table 2.

We then matched our entire sample to the Green Bank northern sky survey at 4.85 GHz (GB6; \citealt{greg96}). The GB6 catalog\footnote{http://www.cv.nrao.edu/$\tilde{\,\,\,\,}$jcondon/gb6ftp.html} includes about 75,000 sources with flux densities above 18 mJy and positional accuracy around 10\arcsec. The survey covers the declination band from 0 to $+$75 degrees (and therefore covers 203 of the 291 objects in our sample). We found 16 matches to type II AGNs in the GB6 catalog within 60\arcsec. The catalog gives peak fluxes which we then converted to integrated fluxes using the cataloged model source extents. For objects with GB6 matches, a reference to \citet{greg96} is given in the last columns of Tables 1 and 2. 

We also matched the sample to the Westerbork Northern Sky Survey at 325 MHz (WENSS; \citealt{reng97}) and the Westerbork in the Southern Hemisphere sky survey at 352 MHz (WISH; \citealt{debr02}). The catalogs compiled from these surveys\footnote{http://www.strw.leidenuniv.nl/wenss/} include more than 300,000 sources to a limiting flux density of approximately 18 mJy. These surveys cover the declination ranges of $\delta>28^o$ (WENSS) and $-26^o<\delta<-9^o$ (WISH), and only 96 type II AGNs from our sample have declinations in these ranges (90 in the WENSS range and 6 in the WISH range). We found 13 matches in the WENSS catalog within 60\arcsec\ of the optical position and none in the WISH catalog. For objects with WENSS matches a reference to \citet{reng97} is given in the last column of Tables 1 and 2.

We also examined the NASA/IPAC Extragalactic Database\footnote{NED, http://nedwww.ipac.caltech.edu/} entries for all the objects in our sample. Those objects that have radio data in the NED other than from GB6, WENSS, FIRST and NVSS are indicated in Tables 1 and 2 with a reference to the NED in the last column. In particular, almost all objects detected by GB6 were also detected by the Texas survey of radio sources at 365 MHz \citep{doug96}. 

All objects with matches at frequencies other than 1.4 GHz are also detected at 1.4 GHz by FIRST and/or NVSS at a level of a few tens of mJy, with no other bright sources in the vicinity, and we therefore consider all these matches to be real counterparts of our sources. Multi-frequency data from GB6, WENSS or other surveys allowed us to determine the radio spectral index $\alpha$ ($F_{\nu}\propto \nu^{\alpha}$) as given in Tables 1 and 2 for 26 objects. If all radio observations cannot be reliably fit with a single power-law, we list the average of the spectral indices on both sides of 1.4 GHz. The calculated index may be affected by fainter nearby sources; due to low resolution of radio data at frequencies other than 1.4 GHz (typically, several arcmin) this effect cannot be taken into account, and we therefore only list spectral indices to one decimal place. For all but one object the calculated spectral indices lie between $-$1 and 0, with most spectral indices $<-0.5$.

\subsection{Radio loudness}
\label{radio_loud}

There is a long-standing discussion in the literature about the existence of the so-called radio-dichotomy, i.e. whether or not the distribution of the radio-to-optical flux ratios of AGNs is bimodal. Some authors argue against the existence of the dichotomy (most recently, \citealt{whit00, cira03}) and some argue in favor (most recently, \citealt{xu99, ivez02, ivez04}). Regardless of the resolution of this controversy, objects on the high end of the radio-to-optical distribution have traditionally been called radio-loud, with the rest being radio-quiet. The fraction of radio-loud objects in the AGN population depends on how the radio-to-optical ratio is defined, but typically 8$-$20\% of optically selected type I AGNs are radio loud \citep{kell89, pado93, hoop96, ivez02}. 

Radio loudness can be defined in terms of the ratio of radio flux to broad-band optical flux, e.g. $B$-band as in \citet{kell89} and \citet{urry95}, or $i$-band as in \citet{ivez02}. Alternatively, radio loudness can be defined in terms of the ratio of the radio luminosity to the narrow emission line luminosity \citep{baum89, xu99}. We adopt the latter approach because broad-band optical fluxes are by definition poor indicators of the intrinsic luminosities of obscured AGNs. Following Paper I, we use the [OIII]5007 emission line as such an indicator.

Radiation observed at 1.4 GHz was emitted at different wavelengths by objects at different redshifts, and we correct the observed fluxes to a common frequency of 1.4 GHz (K-correct). If the observed radio spectrum of an object at redshift $z$ is a power-law, 
\begin{equation}
F_{\nu}=A\nu^{\alpha},
\end{equation}
then the emitted monochromatic luminosity at the same frequency $\nu$ is 
\begin{equation}
L_{\nu}=4\pi D_L^2 F_{\nu} (1+z)^{-1-\alpha}=4\pi D_L^2 F^*_{\nu}.\label{Kcorr}
\end{equation}
Here $D_L$ is the luminosity distance, $(1+z)^{-1-\alpha}$ is the K-correction factor and $F^*_{\nu}$ is the K-corrected flux. Although beaming factors can be important \citep{urry95}, there are not enough observational data to take them into account, and we compute radio luminosities assuming uniform emission into $4\pi$ steradians. Typical spectral indices at a few GHz are in the range $-1<\alpha <0$, for both radio-loud and radio-quiet AGNs \citep{barv89, urry95, ivez04}, as well as for the 26 objects in our sample with multi-frequency radio observations.

\begin{figure}
\epsscale{1.0}
\plotone{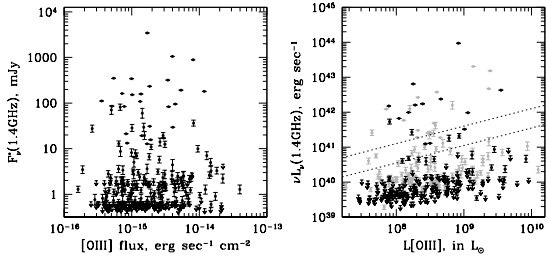}
\figcaption{Radio-optical flux-flux and luminosity-luminosity diagrams at rest-frame 1.4 GHz. The error bars reflect the uncertainty in the spectral index ($-1<\alpha<0$) used in K-correcting. For 26 objects with multi-frequency observations, the K-corrections are calculated using the observed spectral index, and the points are plotted without error bars. The dashed lines on the luminosity-luminosity diagram represent the separation line between radio-loud and radio-quiet objects from \citet{xu99} assuming a radio spectral index of $\alpha=-1$ (top line) and $\alpha=0$ (bottom line). Arrows indicate 5$\sigma$ upper limits from the FIRST survey or 2.3 mJy upper limit from the NVSS survey, K-corrected using $\alpha=-0.5$. On the right diagram, the 179 sources that were selected for spectroscopy only based on their optical properties are indicated by black and the remaining points are grey. \label{lum_lum}}
\end{figure}

Figure \ref{lum_lum} shows radio fluxes vs optical emission line fluxes (left) and luminosities (right, all points) for the entire sample at 1.4 GHz using FIRST and NVSS fluxes and upper limits. The dashed lines on the luminosity-luminosity diagrams represent the separation line between radio-loud and radio-quiet objects from \citet{xu99}, assuming $\alpha=-1$ (top line) and $\alpha=0$ (bottom line). The diagrams for core fluxes and luminosities (not shown) look very similar, since core fluxes differ from total fluxes for only a small number of objects.

As described in Paper I, many objects in the sample were originally targeted for spectroscopy by the SDSS pipeline as counterparts to the FIRST sources. In order to determine the radio-loud fraction, we selected the 179 objects that were targeted for spectroscopy only based on their optical colors or as extended optical sources. These objects are shown in black in Figure \ref{lum_lum} right. We then postulated that the radio properties are not related to any of the optical properties used in the selection of this subsample. Our assumption finds support in a recent study of a large optically selected sample of AGNs by \citet{ivez02}, who showed that the fraction of radio-loud objects does not depend on the AGN luminosity and does not evolve with redshift. However, other studies (e.g., \citealt{pado93}) found that the radio-loud fraction is a function of the optical luminosity. In addition, \citet{ivez02} found that radio-loud AGNs have slightly redder optical colors than radio-quiet AGNs. Therefore, subtle selection effects might affect our estimate of the radio-loud fraction.

Among the 179 optically selected type II AGNs, 12 objects lie above the upper line and 7 more objects lie between the lines. The number of radio-loud objects is then $15.5\pm 3.5$, and the radio-loud fraction is $9\pm 2$\%. Comments on three interesting radio-loud sources can be found in Appendices \ref{SDSS0909}-\ref{SDSS1501}. The radio-loud fraction of type II AGNs is similar to that of broad-line AGNs selected based on their blue colors, and the radio morphologies of the extended sources are similar for type II AGNs and for radio-loud quasars. These similarities suggest that there is no difference between the radio properties of type I and type II AGNs, in agreement with the unification model.

\section{Mid-infrared: IRAS observations}
\label{iras}

The full-sky coverage of the Infrared Astronomical Satellite (IRAS) mission \citep{neug84} makes it invaluable for studies of the mid-infrared sky. IRAS mapped the entire sky in four bands centered at 12, 25, 60 and 100\micron, with a positional uncertainty ellipse for faint sources of $45\arcsec\times (15-20)\arcsec$ (3$\sigma$ at 60 and 100 \micron). 

\subsection{Individual detections}

Fields of 284 type II AGNs from our sample were observed by IRAS; the remaining seven objects happened to lie in the IRAS gaps. We have combined all available IRAS survey observations for each of the 284 objects. Typically, each observation has flux noise of about 100 mJy at 12, 25, and 60\micron\ and 200 mJy at 100\micron, and about 5-15 scans are available for each field, so combining all available observations can result in a factor of 2-4 noise reduction. The data were processed using the routine SCANPI (\citealt{helo88}, Section IIIa), which is available in the NASA/IPAC Infrared Science Archive\footnote{IRSA, http://irsa.ipac.caltech.edu/}. We used median combining of the scans recommended by the on-line SCANPI manual because of the non-Gaussian nature of noise in IRAS data. 

For each object, SCANPI combines all available observations and attempts to fit a point source template at or near the user-specified position (SDSS positions were used). The SCANPI output includes the flux density of the best-fitting template and the offset between the template peak and the input position. The goodness of the fit can be assessed using the template correlation coefficient; for example, a point source detected with a signal-to-noise ratio (S/N) of about 20 typically has a correlation coefficient above 0.995.

SCANPI processing of our sample of type II AGNs resulted in a number of matches, but the template correlation coefficients did not exceed 0.985, and most alleged counterparts are detected at S/N of about 2 or 3. At such low S/N, false matches and noise contribute significantly to the matching rate. To quantify this contamination, we selected a sample of 1600 spectroscopically confirmed stars from the SDSS spectroscopic database with optical magnitudes $17<r<20$. These stars are mostly main sequence stars of stellar types F to M and white dwarfs without any emission lines or other unusual features, and are unlikely to have real IRAS counterparts \citep{knau01}.

\begin{figure}
\epsscale{0.9}
\plotone{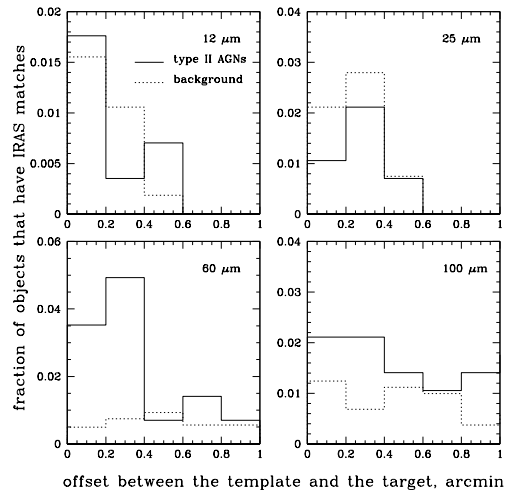}
\figcaption{Stellar counterparts (dotted) and type II AGN counterparts (solid) in IRAS data as found by SCANPI. For this figure, we only used matches that have template correlation coefficients exceeding 0.8. Since the faint stars in this comparison sample are not expected to have any real counterparts, the dotted line can be interpreted as a background estimate. \label{background}}
\end{figure}

Figure \ref{background} shows the matching rate of type II AGNs (solid line). The number of matches with correlation coefficients $>0.8$ is shown as a function of the offset between the peak of the IRAS flux and the optical position. The dotted line shows the number of matches to the stellar sample as an estimate of the background contamination. The matching rate for type II AGNs is roughly the same as the background estimate at 12\micron\ and 25\micron, and we concluded that practically none of the type II AGNs were detected at 12\micron\ or 25\micron. In the 60\micron\ band, 85\% of all matches with correlation coefficients $>$0.8 and offsets $<$0.4\arcmin\ are real IR counterparts of the optical sources, whereas in the 100\micron\ band about half of all matches are background contamination.

Table 3 lists all objects that have possible detections at 60\micron\ and/or 100\micron. For each detection we give a confidence level, which is equal to one minus the probability that the match is a random association, calculated in bins of correlation coefficient and template offset. For example, the highest confidence detections (confidence levels 0.95 in Table 3) are the ones with correlation coefficients above 0.9 and offsets less than 0.4\arcmin. 

\begin{figure}
\epsscale{0.9}
\plotone{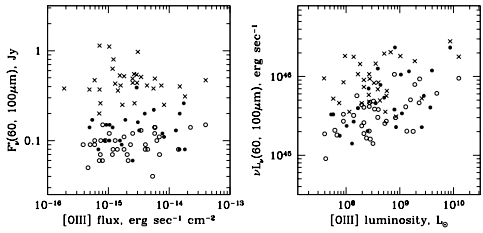}
\figcaption{Mid-IR and optical flux-flux and luminosity-luminosity diagrams: solid circles for high-confidence counterparts at 60\micron, empty circles for low-confidence counterparts at 60\micron, and crosses for counterparts at 100\micron. Spectral index $\alpha=-1$ was assumed, thus no K-corrections are required. \label{oiii_ir}}
\end{figure}

Figure \ref{oiii_ir} shows mid-IR fluxes vs [OIII]5007 line fluxes (left) and luminosities (right). IRAS counterparts at 60\micron\ are classified as high-confidence if the confidence level is 0.8 or greater and as low-confidence otherwise. All 100\micron\ counterparts are low-confidence by this definition. The spectral energy distribution of type II AGNs usually peaks at 50-100\micron, making $\nu L_{\nu}\simeq const$ at these wavelengths \citep{schm97}. Therefore, $\alpha=-1$, and equation (\ref{Kcorr}) implies no K-correction. The directly detected mid-IR luminosity reaches $3\times 10^{46}$ erg sec$^{-1}$ (or $7.5\times 10^{12}L_{\odot}$) for several objects in our sample. 

For most objects, the ratio $\nu L_{\nu}(60, 100\micron)/$ $L$([OIII]5007) is in the range $10^3-10^5$ and is systematically higher than that obtained by \citet{mulc94} for a large sample of nearby AGNs. Since all our detections are very close to the flux limit of the IRAS survey, our IRAS detections are biased toward objects with high IR-to-optical ratios. 

Coadding multiple scans of the same field allows us to significantly increase sensitivity for detecting faint sources. Only three objects in our sample have matches in the IRAS Faint Source Catalog \citep{iras88}: SDSS J005009.81$-$003900.6 (Appendix \ref{SDSS0050}), SDSS J090307.84$+$021152.2, and SDSS J145054.37$+$ 004646.7. Fluxes at 60\micron\ obtained from the SCANPI analysis agree within the errors with those from the Catalog; in addition, we detected SDSS J090307.84$+$021152.2 at 100\micron, whereas the Catalog gives only an upper limit in this band.

\subsection{Averaged mid-IR spectral energy distribution}

The results of the previous section show that IR luminosities of several objects in our sample reach very high values, but our IRAS detections were biased toward objects with high IR-to-optical ratios. In this section we study average IR-to-optical ratios. 

We divided the sample into six redshift bins: four bins with $\Delta z=0.05$ for $0.3<z<0.5$, one bin with $0.5<z<0.6$ and one bin with $0.6<z<0.8$, with each bin containing more than 30 objects. In each bin we median-combined IRAS scans of all objects using the routine SUPERSCANPI which combines observations of different fields. We normalized the obtained IRAS flux by the median [OIII]5007 flux in this bin. This is justified by the observed correlation between the [OIII]5007 line luminosity and the mid-IR luminosity of AGNs \citep{mulc94} and by unification models in which both narrow-line luminosity and the luminosity of the heated obscuring torus scale with the luminosity of the central object. The resulting observed IR-to-optical ratio needs to be corrected to the rest frame. Emission observed at frequency $\nu$ with spectral flux $F_{\nu}$ was emitted at frequency $\bar{\nu}=(1+z)\nu$ and with monochromatic luminosity
\begin{equation}
L_{\bar{\nu}}=4\pi D_L^2 F_{\nu}/(1+z).
\end{equation}
Redshift corrections are small within each redshift bin. This equation is different from equation (\ref{Kcorr}) which relates $L_{\nu}$ and $F_{\nu}$ at the same frequency $\nu$ using assumptions about the spectral shape.

\begin{figure}
\epsscale{1.0}
\plotone{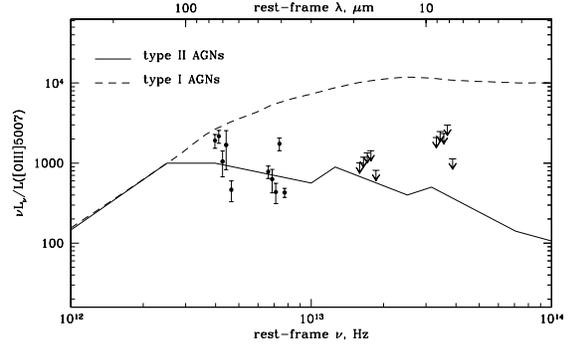}
\figcaption{Average spectral energy distribution of type II AGNs (solid circles and arrows). Our sample of type II AGNs was binned up into five redshift bins, and IRAS scans were coadded for all objects within each redshift bin. The resulting spectral flux was then corrected to the rest frame and normalized to the [OIII]5007 line luminosity. Error bars reflect the nominal 1$\sigma$ uncertainty, but underestimate the intrinsic spread of the IR-to-[OIII]5007 among different objects which can span two orders of magnitude \citep{mulc94}. 3$\sigma$ upper limits for the non-detection of the coadded images at 12\micron\ and 25\micron\ are shown with arrows. The solid and the dashed lines show typical mid-IR spectral energy distributions of type II AGNs \citep{schm97} and of type I AGNs \citep{elvi94}, respectively. These lines are not normalized to the [OIII]5007 luminosity; the type II AGN spectral energy distribution is arbitrarily scaled to have roughly the observed values at $40-80\micron$, and the type I AGN spectral energy distribution is scaled to coincide with it at long wavelengths. The spectral energy distribution of AGNs is poorly known longward of 100\micron. \label{pic_sed_comparison}}
\end{figure}

The resulting spectral energy distribution is presented in Figure \ref{pic_sed_comparison}. The central source is detected in all redshift bins at 60 and 100\micron\ at greater than 4$\sigma$ level, but is not detected at 12\micron\ or 25\micron, and 3$\sigma$ upper limits are shown for these bands. If we assume that the spectrum between rest-frame 40\micron\ and 80\micron\ is roughly flat ($\nu L_{\nu} \simeq const$), we can obtain the average luminosity in this spectral region:
\begin{equation}
L(40-80\mu{\rm m})\simeq 730 \times L\mbox{([OIII]5007)}.\label{ir-oiii}
\end{equation}
This ratio is very similar to that found by \citet{mulc94} for type I and type II AGNs. Their mean IR/[OIII]5007 ratios are about 400 for type I AGNs and 600 for type II AGNs, and the uncertainty of these values is about 0.8 dex. 

In Paper I we showed that luminous quasars ($M_B<-23$) typically have [OIII]5007 line luminosities in excess of $3\times 10^8 L_{\odot}$ and hence we set this value as a criterion to distinguish between type II quasars and Seyfert 2 galaxies. At this [OIII]5007 luminosity, $L(40-80\micron)\simeq 0.9\times 10^{45}$ erg sec$^{-1}$. This is still only a lower bound on the bolometric luminosity, and significant energy can be emitted at longer and shorter wavelengths in the IR. The upper limits at 12\micron\ and 25\micron\ are consistent with the typical spectral energy distribution of type II AGNs (\citealt{schm97}, solid line in Figure \ref{pic_sed_comparison}) and rule out type I spectral energy distributions in which $\nu L_{\nu}$ rises by an order of magnitude toward 10\micron\ \citep{elvi94}. We have verified that the results of the SUPERSCANPI analysis are insensitive to whether or not the sources with high-confidence individual detections are included in the coaddition. 

Although the scatter of points associated with different redshift bins is large, IRAS data give an indication that the spectral energy distribution is approximately flat ($\nu L_{\nu}\simeq const$) from 40 to 70\micron\ in the rest-frame (Figure \ref{pic_sed_comparison}). This suggests that emission from cool dust (temperatures of a few tens of K) can make a significant contribution to the bolometric luminosity of type II AGNs. Such a cool component is common in both type I and type II AGNs (e.g., \citealt{degr92, kura03}), but its origin is not well understood. It can be reproduced using models that involve two-component heating (from the AGN and from the starburst, e.g., \citealt{rowa89}) or in some models of the AGN dust disk (e.g., the warped disk model by \citealt{sand89} and the clumpy disk model by \citealt{nenk02}).

\subsection{Comparison with broad-line AGNs}
\label{sec_compare}

\begin{figure}
\epsscale{1.0}
\plotone{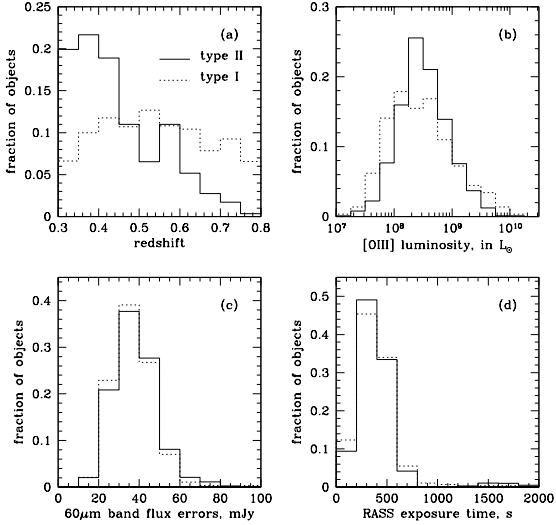}
\figcaption{(a) redshift distribution, (b) [OIII]5007 line luminosity distribution, (c) IRAS 60\micron\ band spectral flux error distribution, and (d) RASS exposure time distribution for type I AGNs (dotted line) and type II AGNs (solid line).\label{pic_comp_distrib}}
\end{figure}

In order to make a direct comparison between the IR properties of type II AGNs and type I AGNs, we selected 1784 type I AGNs from 300 spectroscopic plates from the publicly available SDSS Data Release 1 \citep{abaz03} in the redshift range $0.3<z<0.8$ with the following selection criteria: (a) the continuum is dominated by the AGN, not by the host galaxy; (b) there is at least one permitted emission line in the spectrum with width larger than that of the forbidden line [OIII]5007; (c) the optical spectrum does not show significant reddening or obscuration. Specifically, we rejected AGNs with red spectra ($f_{\lambda}$ increasing with $\lambda$) and objects with low continuum and a broad component in H$\alpha$ but not in H$\beta$ (so-called type 1.9 AGNs). The redshift and [OIII]5007 distributions of this comparison sample are shown in Figures \ref{pic_comp_distrib}a and b. Figure \ref{pic_comp_distrib}c shows that the IRAS flux limits are the same for the sample of type II AGNs and for the comparison sample of type I AGNs. Although nominal flux errors underestimate the real error, for each field they are calculated using the number of scans of this field and therefore represent an adequate measure of the IRAS depth at a given position. 

\begin{figure}
\epsscale{1.0}
\plotone{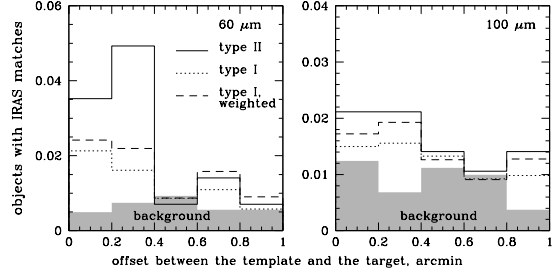}
\figcaption{IRAS detection rates of type II AGNs (solid line) and type I AGNs (dotted and dashed lines) in the 60\micron\ and 100\micron\ bands. For this figure, we only used matches that have template correlation coefficients exceeding 0.8. Detections of type I AGNs (dotted line) were weighted according to the positions of the detected AGNs in the redshift-[OIII]5007 luminosity plane (dashed line). The weighted detection rate of type I AGNs can be directly compared to the detection rate of type II AGNs. The background estimate (shaded histogram) is the same as in Figure \ref{background}. \label{pic_iras_comparison}}
\end{figure}

The IRAS detection rates of type I and type II AGNs are shown in Figure \ref{pic_iras_comparison}, but they cannot be compared directly because the comparison sample of type I AGNs is, on average, farther away because type I AGNs are brighter in the optical and can be seen out to higher redshifts (Figure \ref{pic_comp_distrib}a). Furthermore, the [OIII]5007 emission line contributes significantly to the broad-band flux in type II AGNs (Paper I) and can alone push an otherwise faint object above the SDSS spectroscopic flux limit, so type II AGNs are slightly more luminous than type I AGNs in the [OIII]5007 emission line (Figure \ref{pic_comp_distrib}b). 

To address this difference statistically, one could select a subsample of type I AGNs with the same redshift and [OIII]5007 luminosity distributions as the type II AGNs from our sample. However, there are simply too few luminous type I AGNs at low redshifts in the SDSS spectroscopic samples. Instead, we adopted a technique to weight each type I AGN in the comparison sample according to its redshift and [OIII]5007 luminosity so that the comparison sample becomes directly comparable to the type II AGN sample.

We binned the redshift-$L$[OIII]5007 plane in 30 bins (six bins in redshift times five bins in $L$([OIII]5007)) and calculated the number of type I AGNs per each bin ($m_i$, $\sum_i m_i=N_I=1784$) and the number of type II AGNs per each bin ($n_i$, $\sum_i n_i=N_{II}=284$, the seven objects without IRAS coverage excluded). To each type I AGN we then assigned a weight 
\begin{equation}
w=\frac{n_i}{N_{II}}\cdot\frac{N_I}{m_i}\label{weight}
\end{equation}
depending on its redshift and [OIII]5007 luminosity. Then, by definition, the weighted redshift and [OIII]5007 luminosity distributions (separately for each redshift bin and for the whole sample) of type I AGNs are identical to those of type II AGNs. Clearly, nearby type I AGNs with high [OIII]5007 luminosity received the highest weights. 

The weighted detection rates of type I AGNs are shown with a dashed line in Figure \ref{pic_iras_comparison} and can be compared to the detection rates of type II AGNs. They are similar for the 100\micron\ band, but in the 60\micron\ band the detection rate of type II AGNs appears to be about twice as high as that of type I AGNs. The difference in the 60\micron\ band is statistically significant (the Poisson probability of the observed detection rate with the average given by the type I AGN detection rate is less than 4\%). Based on the narrow line ratios of the composite spectrum of type II AGNs, it was suggested in Paper I that the average reddening of the Balmer-emitting region is about $E(B-V)=0.27$ mag. If this is also true of the [OIII]5007 emitting region, then the average extinction of the [OIII]5007 line is a factor of 2.3. This would imply that the type II AGNs from our sample are on average about twice as luminous in the [OIII]5007 line as type I AGNs. Such difference in the intrinsic luminosity would account for the difference in the detection rate. This result is similar to that of \citet{mulc94}, who found that the IR-to-[OIII]5007 ratios are slightly higher for type II AGNs than for type I AGNs. The difference of the detection rates in the 100\micron\ band is not statistically significant, and the error in the detection rate is much higher in this band.

\section{Near-infrared: 2MASS observations}
\label{nir}

The Two Micron All Sky Survey (2MASS, \citealt{skru97}) has mapped the entire sky in the $J$ (1.13-1.37\micron), $H$ (1.50-1.80\micron), and $K_S$ (2.00-2.32\micron) photometric bands. The 2MASS Point Source Catalog (PSC, \citealt{cutr03}) includes $\sim 10^8$ objects with limiting magnitudes of 15.8, 15.1 and 14.3 in the three bands, respectively ($10\sigma$ level) and with positions accurate to $<0.5\arcsec$. The PSC also includes extended sources.

We have positionally matched the sample of the 291 type II AGNs from the SDSS to the 2MASS PSC and found 42 matches within 2\arcsec. The median positional offset between the SDSS and the 2MASS coordinates is 0.39\arcsec. Our sample is faint in the optical ($\langle r\rangle=20.4$), and only the brightest objects from our sample were matched. The median optical magnitude of the objects with matches in 2MASS is $\langle r\rangle=19.7$. The interesting question that can be probed with the near-IR data is whether our optically obscured type II AGN candidates remain obscured at near-IR wavelengths. Partially obscured AGNs will appear red in near-IR colors \citep{bark01, hopk04}, and the near-IR colors of reddened AGNs vary with redshift \citep{rich03, fran04}. At the relatively low redshifts of the objects in our sample, about 90\% of optically selected quasars have $J-K_S<2$. AGNs that have $J-K_S>2$ are considered reddened, although this criterion is somewhat arbitrary \citep{fran04}. 

\begin{figure}
\epsscale{1.0}
\plotone{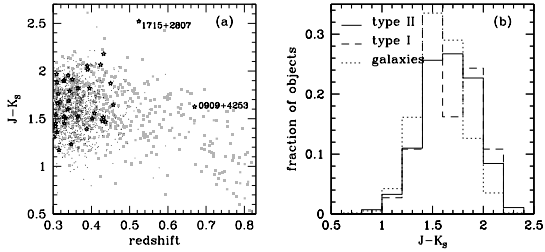}
\figcaption{Left: redshift-color diagram for 2MASS counterparts of type II AGNs (stars), the comparison sample of type I AGNs (grey squares) and galaxies (dots). Right: color distribution of galaxies (dotted line), type I AGNs (dashed line) and type II AGNs (solid line) for objects in the redshift range $0.3<z<0.5$. Near-IR colors of type II AGNs are mostly consistent with being dominated by the light from the host galaxies. The outliers are a red AGN SDSS J171559.79$+$280716.8 \citep{cutr02} and a radio-loud AGN SDSS J090933.51$+$425346.5 with continuum dominated by synchrotron radiation (Appendix \ref{SDSS0909}). \label{pic_zc}}
\end{figure}

Figure \ref{pic_zc}a shows the redshift-color diagram for type II AGNs (stars), type I AGNs from our comparison sample (grey squares) and galaxies (black dots), and Figure \ref{pic_zc}b compares the color distributions of the three samples. If the nuclear light of type II AGNs from our sample was detected in the near-IR, they would appear much redder than type I AGNs, typically with $J-K_S>2$ \citep{cutr02, fran04}. With the exception of a few objects this is not the case, as can be seen from Figure \ref{pic_zc}, and the near-IR colors of type II AGNs are consistent with being dominated by the light from the host galaxies. Galaxies and type I AGNs at low redshifts coincidentally have similar red $J-K_S$ color (Figure \ref{pic_zc}); for type I AGNs the red color is due to the near-IR bump in the spectral energy distribution \citep{sand89}. As this bump is redshifted out of the $K_S$ band, the colors of type I AGNs become bluer. 

\section{X-rays: ROSAT observations}
\label{x}

The ROSAT All-Sky Survey (RASS, \citealt{voge99, voge00}) imaged the entire sky in the 0.1$-$2.4 keV range with a typical limiting sensitivity for point sources of 10$^{-13}$ erg sec$^{-1}$ cm$^{-2}$. The exact sensitivity limit depends on the position on the sky, because exposure times varied systematically as a function of ecliptic latitude and with a significant variation between scans \citep{voge99}. About 10$^5$ sources are contained in the RASS Bright and Faint Source Catalogs, with a positional uncertainty of 10-30\arcsec. 

\subsection{Matching to RASS catalogs and the detection rate}

We used the database of the High Energy Astrophysics Science Archive Research Center\footnote{HEASARC, http://heasarc.gsfc.nasa.gov/} to match the sample of type II AGNs to the RASS Bright and Faint Source Catalogs. The matching radius was chosen to be 1\arcmin, because chance associations start dominating on angular scales in excess of 1\arcmin, and 90\% of matches within this radius are real counterparts \citep{voge99, ande03}. We found six matches in RASS catalogs within 1\arcmin\ of the optical positions of type II AGNs (one match in the RASS Bright Source Catalog and five matches in the Faint Source Catalog). The six matches and their X-ray properties are listed in Table 4. 

Only six out of 291 sources from our sample (2\%) have RASS counterparts; this fraction of RASS sources is much lower than in other low-redshift SDSS AGN samples. In order to quantify this difference, we used the comparison sample of 1784 type I AGNs described in Section \ref{sec_compare}. The distribution of RASS exposure times for the two samples is very similar (Figure \ref{pic_comp_distrib}d). Among the 1784 type I AGNs from the comparison sample, 273 objects (15\%) have matches in the RASS catalogs within 1\arcmin\ of the optical position. Again, for proper comparison of detection rates we weight the detections of type I AGNs to the procedure described in Section \ref{sec_compare}; we find that the weighted detection rate of type I AGNs is about 21\%. 

Thus type II AGNs are ten times less likely to be detected by RASS than are type I AGNs with the same redshift and [OIII]5007 distribution. This implies that the soft X-ray properties of the objects from our sample are significantly different from those of type I AGNs. A natural explanation for this difference is that soft X-rays are absorbed along the line of sight. A hydrogen column density of $N_H=2\times 10^{22}$ cm$^{-2}$ suppresses observed counts in the ROSAT range (0.1$-$2.4 keV) by a factor of ten. If type II AGNs have intrinsic luminosities similar to those of the comparison type I AGNs, such column densities would produce roughly the observed difference in the detection rates. An alternative explanation is that type II AGNs are intrinsically X-ray faint. We are conducting follow-up observations of the objects in our sample with $Chandra$ and $XMM$ to distinguish between these two possibilities and to determine intrinsic X-ray luminosities and column densities along the line of sight. 

\subsection{Properties of RASS-detected type II AGNs}

\begin{figure}
\epsscale{1.0}
\plotone{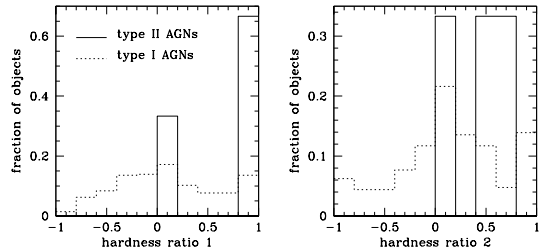}
\figcaption{Distributions of the RASS hardness ratios for 173 RASS-detected type I AGNs (dotted histogram) and six RASS-detected type II AGNs (solid histogram) with RASS detections. Distributions of hardness ratios of type I and type II AGNs are statistically different at the 97\% KS confidence level (for HR1) and at the 90\% KS confidence level (for HR2). \label{pic_hardness}}
\end{figure}

For the objects with RASS detections, two hardness ratios are available in the RASS catalogs. Hardness ratios HR1 and HR2 are defined in \citet{voge99}; both values can vary between $-1$ and 1. Objects with harder spectra have higher hardness ratios. In Figure (\ref{pic_hardness}) we compare the distributions of the hardness ratios of type I and type II AGNs. The errors of individual hardness ratios are too large to use them to study the spectral shape of individual objects, but on average both hardness ratios seem to be higher in the six RASS-detected type II AGNs than in RASS-detected type I AGNs. Using the Kolmogorov-Smirnov test, we found that hardness ratios of type II AGNs and type I AGNs represent different populations at 97\% confidence level (for HR1) and at 90\% confidence level (for HR2). This result should be treated with caution as hardness ratios of type II AGNs are highly uncertain due to low RASS fluxes, and only six objects were detected which are not necessarily representative of the entire sample. We will be able to study spectral properties of type II AGNs in much greater detail using $Chandra$ and $XMM$ data. 

In Table 4, together with the observed RASS counts and hardness ratios, we list X-ray luminosities in the rest-frame 0.1$-$2.4 keV range assuming a power-law spectrum with photon index $\Gamma=0$ and corrected for the Galactic column density\footnote{Calculated using WebPIMMs tool available online at HEASARC.}. Although in hard X-rays (2$-$10 keV) typical photon indices of AGNs are in the range 1.5$-$2.5 (e.g., \citealt{bran97}), in many objects intrinsic absorption and reprocessing result in a flatter observed spectral energy distribution in the soft band, and we use $\Gamma=0$ to account for this effect. Since our calculation does not correct for any intrinsic obscuration, it places a conservative lower bound on the intrinsic X-ray luminosity. All six sources have observed X-ray luminosities around or above $10^{44}$ erg sec$^{-1}$ and would be classified as quasars based on their X-ray luminosity (e.g., \citealt{szok04}). In the optical, all but one object have [OIII]5007 luminosity around or above the quasar limit of $3\times 10^8 L_{\odot}$. 

Three out of six RASS type II AGNs are radio-loud (Table 4), whereas only one would be expected to be radio-loud if six objects were randomly drawn from our sample. Because of the limited angular resolution and flux sensitivity of previous-generation X-ray missions not many X-ray counterparts to radio jets are known, but $Chandra$ has been finding an increasing number of such objects \citep{samb02}. In some cases observed X-ray luminosities of emission associated with radio jets reach $10^{44}-10^{45}$ erg sec$^{-1}$ (e.g., \citealt{fabi03, scha03}). One intriguing possibility is that some of the X-ray emission in the three radio-loud objects detected by ROSAT could be associated with their jet activity. 

\section{Conclusions}
\label{conclusions}

In this paper we have studied multi-wavelength properties of 291 type II AGNs at redshifts $0.3<z<0.8$ selected from the SDSS spectroscopic database based on their emission line properties. The details of the selection and optical properties of the sample are described in Paper I \citep{z03}. In particular, we showed that about 50\% of the objects in our sample (those that have [OIII]5007 luminosities in excess of $3\times 10^8 L_{\odot}$) are luminous enough to be classified as type II quasars.

Of the type II AGNs in our sample, we found 143 matches in the FIRST survey (20 cm). This high fraction of FIRST detections is due to specific targeting of FIRST sources by the SDSS spectroscopic survey. When this selection effect is accounted for, the fraction of the radio-loud objects in our sample is found to be $9\pm 2$\%. This is consistent with estimates of the radio-loud fraction of the AGN population as a whole. The radio properties of type II AGNs from our sample (radio morphologies, spectral indices) are similar to those of type I AGNs. These conclusions are in agreement with unification model of AGNs, since radio emission is not affected by circumnuclear obscuration.

Most objects in the sample are fainter than the flux limit of the IRAS Point Source Catalog, but coaddition of all available IRAS scans of our sources yielded several $\sigma$ detections (flux levels of 50$-$250 mJy) in the 60\micron\ and/or 100\micron\ IRAS bands for about 40 objects in our sample. The inferred infrared luminosities of these objects are in the range $\nu L_{\nu}(60-100\mu{\rm m})=10^{45}-3\times 10^{46}$ erg sec$^{-1}=2.5\times 10^{11}-7.5 \times 10^{12} L_{\odot}$, placing them among the most luminous quasars known at the redshifts of our sample ($0.3<z<0.8$). Possible confusion with other sources and low sensitivity place severe limits on the amount of information that can be deduced from the IRAS data. In particular, we cannot say anything about the shapes of the IR spectral energy distributions of individual objects, and although the inferred $\nu L_{\nu}$ places a very interesting lower bound on the bolometric luminosity, the latter is still largely unknown. 

Coaddition of IRAS data for all the objects in our sample demonstrates that the sample as a whole has a large IR-to-[OIII]5007 ratio, similar to that of other AGNs. The coadded spectral energy distribution $\nu L_{\nu}$ is roughly constant as a function of $\nu$; although the errors of $\nu L_{\nu}$ are large, $\nu L_{\nu}$ is approximately constant from 40 to 70\micron\ in the rest-frame suggesting that emission from cool material (a few tens of K) is energetically important. We estimate that the median IR luminosity of the objects in our sample is $\nu L_{\nu}(40-80\micron)=0.9\times 10^{45}$ erg sec$^{-1}$ (corresponding to the median $L$([OIII]5007)$=3\times 10^8 L_{\odot}$).

Only six out of 291 type II AGNs are detected by RASS in the soft X-rays (0.1$-$2.4 keV). This detection rate is about ten times less than for type I AGNs with the same redshift and [OIII]5007 luminosity distribution. A column density of $2\times 10^{22}$ cm$^{-2}$ applied to type I AGNs X-ray spectra would suppress emission in the range 1$-$2 keV by a factor of ten and produce the observed difference in the detection rates. 

Data from large surveys covering a large range of wavelengths have provided very interesting constraints on the spectral energy distributions of the objects in our sample. We are conducting sensitive pointed follow-up observations of type II AGNs with $Spitzer$, $Chandra$ and $XMM$ telescopes to determine the details of the spectral energy distributions. This will allow us to determine bolometric luminosities and constrain models of circumnuclear obscuration.  

\section*{Acknowledgments}

The authors would like to thank Iskra Strateva for helpful suggestions and detailed comments on the manuscript. We are also grateful to the referee for important suggestions that significantly improved the paper.

Funding for the creation and distribution of the SDSS Archive has been provided by the Alfred P. Sloan Foundation, the Participating Institutions, the National Aeronautics and Space Administration, the National Science Foundation, the U.S. Department of Energy, the Japanese Monbukagakusho, and the Max Planck Society. The SDSS Web site is http://www.sdss.org/. 

The SDSS is managed by the Astrophysical Research Consortium (ARC) for the Participating Institutions. The Participating Institutions are The University of Chicago, Fermilab, the Institute for Advanced Study, the Japan Participation Group, The Johns Hopkins University, Los Alamos National Laboratory, the Max-Planck-Institute for Astronomy (MPIA), the Max-Planck-Institute for Astrophysics (MPA), New Mexico State University, University of Pittsburgh, Princeton University,
 the United States Naval Observatory, and the University of Washington.

This research has made use of the NASA/IPAC Extragalactic Database (NED, \\http://nedwww.ipac.caltech.edu/) and the \\NASA/IPAC Infrared Science Archive (IRSA, \\http://irsa.ipac.caltech.edu/) which are operated by the Jet Propulsion Laboratory, California Institute of Technology, under contract with the National Aeronautics and Space Administration. The authors would like to thank John Good (IRSA) for technical support. This research has made use of the High Energy Astrophysics Science Archive Research Center (HEASARC, http://heasarc.gsfc.nasa.gov/) operated by the Laboratory for High Energy Astrophysics at NASA/GSFC and the High Energy Astrophysics Division of the Smithsonian Astrophysical Observatory. 

NLZ and MAS acknowledge the support NSF grant AST-0307409 and HST grant HST-GO-09905.01.

\appendix
\section{Previously known and unusual objects}
\label{notes}

In this section we briefly summarize details of multi-wavelength analysis for two previously known objects (SDSS J005009.81$-$003900.6 and SDSS J090933.51$+$425346.5) that had been classified as an Ultraluminous Infrared Galaxy and a blazar, correspondingly. We also discuss radio morphology of the known radio source SDSS J121637.25$+$672441.5 and comment on the unusual radio morphology of SDSS J150117.96$+$545518.3. 

\subsection{SDSS J005009.81$-$003900.6}
\label{SDSS0050}

\begin{figure}
\epsscale{1.0}
\plotone{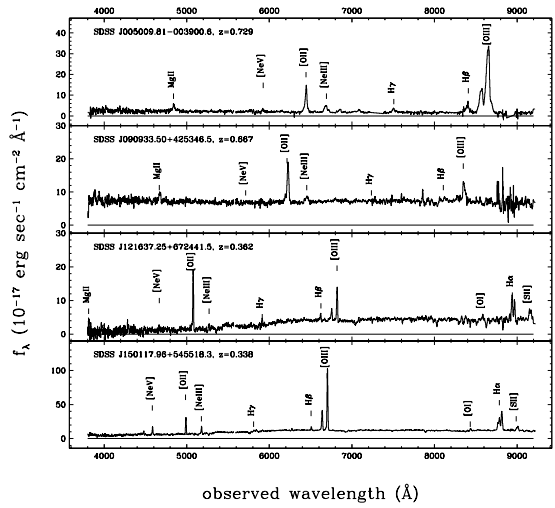}
\figcaption{Optical spectra of four type II AGN candidates described in the Appendices. Spectra are smoothed by 5 pixels, the thin line is at the zero flux density level. \label{pic_appendix}}
\end{figure}

This object is a known Ultraluminous Infrared Galaxy at redshift $z=0.729$ with IR luminosity $\nu L_{\nu}=2.8\times 10^{46}$ erg sec$^{-1}$ at rest-frame 35-60\micron\ and with $L$([OIII]5007)$=9\times 10^9 L_{\odot}$. It was selected by \citet{stan00} by comparing the IRAS Faint Source Catalog with the FIRST Point Source Catalog. The radio flux is 4.3mJy, and therefore the source is radio-quiet. Although most of the objects in the sample by \citet{stan00} are classified as starburst galaxies, this object shows high-ionization emission lines [NeV]3346,3426 and a high [OIII]5007/H$\beta$ ratio characteristic of AGN activity (Figure \ref{pic_appendix} top). 

\subsection{SDSS J090933.51$+$425346.5}
\label{SDSS0909}

This object (also known as 3C216) is a radio-loud quasar with $L$([OIII]5007)$=8.3\times 10^8L_{\odot}$ at redshift $z=0.667$. High-resolution radio observations indicate that there are apparent superluminal motions present in the radio core of this object, and the object has been classified as a blazar \citep{bart88}. The radio morphology shows rich substructure on sub-arcsec scales, with a prominent jet that bends sharply producing a bright knot \citep{feje92}. The FIRST image is resolved (Figure \ref{extended} left column, middle). The optical spectrum (Figure \ref{pic_appendix}) shows bright narrow emission lines. Contrary to most other type II quasar candidates, it shows a bright optical continuum with $f_{\lambda}\simeq const$ which has no Balmer break or strong absorption lines and therefore cannot be attributed to the host galaxy. Most probably, it is synchrotron radiation from the radio jet (the optical continuum spectrum is typical of BL Lac objects). Another peculiar feature of this object is that $L$([OII]3727)$\simeq 2L$([OIII]5007), whereas in the majority of type II AGNs [OIII]5007 is the most luminous emission line of the optical spectrum. 

\subsection{SDSS J121637.25$+$672441.5}
\label{SDSS1216}

The NVSS image of the field of SDSS J121637.25$+$672441.5 shows a double-lobed structure with a peak separation of about 4.6\arcmin\, but no prominent core component can be identified. As a large angular size radio source, this object was followed up with high-resolution radio mapping using VLA by \citet{lara01a}. These observations allowed the authors to identify the core component, for which they later found an optical counterpart coincident with SDSS J121637.25$+$672441.5 and presented its optical spectroscopy \citep{lara01b}. The maximum extent of the radio structure seen on the VLA image is about 6\arcmin, which at the redshift of the object $z=0.362$ corresponds to 1.9 Mpc, placing it among the largest radio structures known. In the optical, it displays a typical Seyfert 2 spectrum (Figure \ref{pic_appendix}) with an [OIII]5007 luminosity of $1.5\times 10^8 L_{\odot}$, below the quasar limit. 

\subsection{SDSS J150117.96$+$545518.3}
\label{SDSS1501}

\begin{figure}
\epsscale{1.0}
\plotone{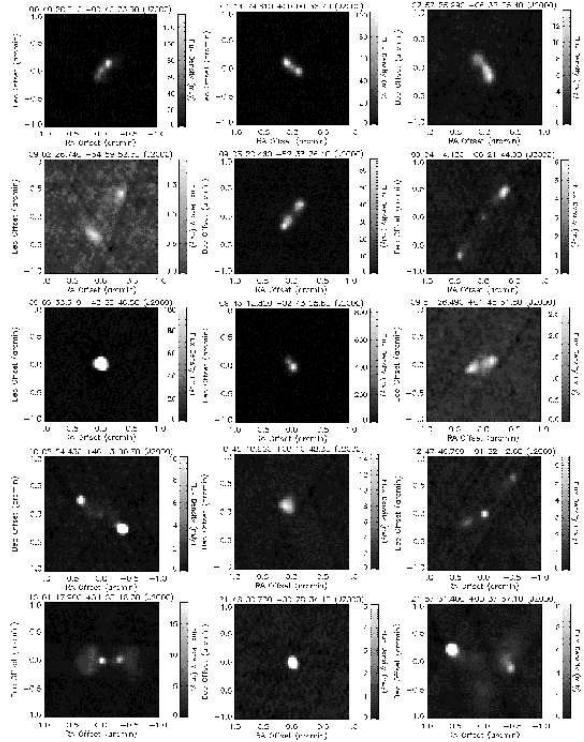}
\figcaption{Extended FIRST matches to type II AGNs (two more extended objects are shown in Figure \ref{confidence-matches}). The size of all stamps is 2\arcmin$\times$2\arcmin; they are centered on the SDSS positions (given on top of each stamp); the vertical bar on the right gives the greyscale code for brightness. The brightness scale was chosen to highlight the morphology, and one or more components are saturated on the chosen brightness scale for SDSS J$0909+4253$, SDSS J$1008+4613$, SDSS J$2148-0028$ and SDSS J$2157+0037$. [The Figure is given in low resolution because of the size restrictions of the Archive. For a high-resolution version, see http://www.astro.princeton.edu/$\tilde{\,\,\,}$nadia/ qso2.html.] \label{extended}}
\end{figure}

This object is a radio-loud type II quasar candidate at $z=0.338$ and with $L$([OIII]5007)$=1.1\times 10^9 L_{\odot}$. Its radio morphology is shown in Figure \ref{extended} bottom left. It has a symmetric shape with two bright knots separated by 13\arcsec\ with flux ratio of about 1.9. Interestingly, the optical counterpart is not centered between the two radio components, but rather coincides with the brighter of the two knots (the radio stamp in Figure \ref{extended} is centered on the optical position). This object also has a luminous RASS counterpart (see Table 4). It is unlikely that SDSS J150117.96$+$545518.3 is just a positional coincidence with the radio source, since there are no other objects in the SDSS field within 17\arcsec\ of SDSS J150117.96$+$545518.3 down to a limiting magnitude of $r=22.2$ that could be responsible for the radio source.

\clearpage

\clearpage
\begin{deluxetable}{cccccccc}
\tabletypesize{\tiny}
\tablewidth{0pt}
\setlength{\tabcolsep}{0.03in}
\tablecaption{FIRST matches of type II AGNs\label{table_first_1}}
\tablehead{&redshift, & &$F_{\nu}$(1.4 GHz), &$F_{\nu}$(1.4 GHz), &spectral & &\\
J2000 coordinates &z & log($L$([OIII]5007)/$L_{\odot}$) & mJy (core) & mJy (total) & index & morphology & references}
\startdata 
SDSS J093818.57$+$005826.8 &0.493 & 8.09 &   0.9 &   0.9 &     &1 &\\ 
SDSS J094209.00$+$570019.7 &0.350 & 8.31 &   0.9 &   0.9 &     &1 &\\ 
SDSS J094312.82$+$024325.8 &0.592 & 9.14 &   0.0 &1331.5 &-0.5 &2 &(1,2,4)\\ 
SDSS J094557.03$+$570803.2 &0.512 & 8.32 &  25.2 &  25.2 &-0.6 &1 &(2,3,4)\\ 
SDSS J094820.38$+$582526.6 &0.353 & 7.89 &   2.4 &   2.4 &     &1 & 
\enddata

\tablecomments{This table includes all matches with the FIRST catalog. J2000 coordinates, redshifts and the [OIII]5007 luminosities (given as log($L$([OIII]5007)/$L_{\odot}$)) were taken from Paper I. Core and total fluxes are identical for point sources. For extended sources, core fluxes are set to 0 if no core component is identified. Some total fluxes for extended sources were taken from the NVSS catalog \citep{cond98} in which case reference (1) is listed in the last column. Spectral indices are available for 23 sources with multi-wavelength radio observations. The radio morphology flag is set to 1 for sources with unresolved radio emission and to 2 for sources with extended radio emission. Table 1 is published in its entirety in the electronic edition of the Astronomical Journal. A portion is shown here for guidance regarding its form and content. }
\tablerefs{(1) \citet{cond98}, (2) \citet{greg96}, (3) \citet{reng97}, (4) other (NED)}
\end{deluxetable}

\begin{deluxetable}{cccccccc}
\tabletypesize{\tiny}
\tablewidth{0pt}
\setlength{\tabcolsep}{0.03in}
\tablecaption{Radio matches of type II AGNs not covered by FIRST\label{table_first_2}}
\tablehead{&redshift, & & offset, &$F_{\nu}$(1.4 GHz), &spectral & & \\
J2000 coordinates &z & log($L$([OIII]5007)/$L_{\odot}$) & arcmin & mJy (total) & index & morphology & references}
\startdata 
SDSS J$032939.85+005220.0$ &0.446 &  8.28 & 0.1 & 159.1 &  -0.9 & 1 & (1,2,4)\\ 
SDSS J$033248.50-001012.3$ &0.310 &  8.50 & 0.2 & 38.8 & 0.0 & 2 & (1,4)\\ 
SDSS J$103951.49+643004.2$ &0.402 &  9.41 & 0.1 &  3.9 & & 1 & (1)\\ 
SDSS J$121637.27+672441.6$ &0.362 &  8.18 & 0.0 & 179.9 & -0.6 & 2 & (1,2,3,4)\\ 
SDSS J$211742.59+005708.0$ &0.486 &  8.78 & 0.0 & 17.1 & & 1 & (1) 
\enddata 

\tablecomments{This table includes all matches with radio catalogs within 30\arcsec\ for type II AGNs not observed by FIRST. J2000 coordinates, redshifts and the [OIII]5007 luminosities (given as log($L$([OIII]5007)/$L_{\odot}$)) were taken from Paper I. In the fourth column, offsets between the optical and the radio position are given in arcmin. Morphology is given based on the NVSS images (1 if a point source, 2 if extended). }
\tablerefs{(1) \citet{cond98}, (2) \citet{greg96}, (3) \citet{reng97}, (4) other (NED)}
\end{deluxetable}

\begin{deluxetable}{ccccccccc}
\tabletypesize{\tiny}
\tablewidth{0pt}
\setlength{\tabcolsep}{0.03in}
\tablecaption{IRAS counterparts of type II AGNs\label{table_iras}}
\tablehead{&redshift, & &$F_{\nu}$(60\micron), &$\sigma_{\nu}$(60\micron), & &$F_{\nu}$(100\micron), &$\sigma_{\nu}$(100\micron), & \\
J2000 coordinates &z & log($L$([OIII]5007)/$L_{\odot}$) & mJy & mJy &confidence & mJy & mJy &confidence}
\startdata 
SDSS J$002531.46-104022.2$ &0.303 &  8.73 & & & & 410 & 123 &0.50 \\ 
SDSS J$002852.87-001433.6$ &0.310 &  8.43 & 130 &  37 &0.75 & 510 & 147 &0.50 \\ 
SDSS J$005009.81-003900.6$ &0.729 &  9.94 & 200 &  49 &0.95 & 400 & 106 &0.50 \\ 
SDSS J$005621.72+003235.8$ &0.484 &  9.45 & 130 &  30 &0.80 & & & \\ 
SDSS J$011228.08-010058.2$ &0.388 &  7.98 & & & &1140 & 312 &0.50 
\enddata

\tablecomments{J2000 coordinates, redshifts and the [OIII]5007 luminosities (given as log($L$([OIII]5007)/$L_{\odot}$)) were taken from Paper I. In columns $F_{\nu}$(60, 100\micron) we list spectral fluxes in mJy obtained by fitting a point source template to the sum of the IRAS scans at the position of the object. The nominal spectral flux error is given in columns $\sigma_{\nu}$(60, 100\micron), but it is an underestimate of the real error. The `confidence' columns contain estimated probability that the match is not a random superposition. The seven sources without IRAS coverage are SDSS J104210.95$+$001048.3, SDSS J104807.74$+$005543.4, SDSS J223136.27$-$011045.0, SDSS J223959.04$+$005138.3, SDSS J224409.48$-$083505.2, SDSS J224950.42$+$005157.2, and SDSS J225721.78$-$100000.9. Table 3 is published in its entirety in the electronic edition of the Astronomical Journal. A portion is shown here for guidance regarding its form and content. }
\end{deluxetable}

\begin{deluxetable}{ccccccccc}
\tabletypesize{\tiny}
\tablewidth{0pt}
\setlength{\tabcolsep}{0.03in}
\tablecaption{RASS counterparts of type II AGNs\label{table_rass}}
\tablehead{&redshift, & & offset, &count rate, & & & & radio \\
J2000 coordinates &z & log($L$([OIII]5007)/$L_{\odot}$) & arcmin & counts sec$^{-1}$ & HR1 & HR2 & log($L_X$) & loud? }
\startdata 
SDSS J$011429.61+000036.7$ &0.389 &  8.66 &0.38 &0.029 &1.0 &0.4 &44.24 &yes\\ 
SDSS J$033248.50-001012.3$ &0.310 &  8.50 &0.12 &0.017 &1.0 &0.6 &43.87 &yes\\ 
SDSS J$123215.81+020610.0$ &0.480 &  9.69 &0.75 &0.019 &0.8 &0.2 &44.21 &no\\ 
SDSS J$143047.33+602304.5$ &0.607 &  8.44 &0.44 &0.022 &1.0 &0.5 &44.46 &no\\ 
SDSS J$150117.96+545518.3$ &0.338 &  9.06 &0.12 &0.050 &0.0 &0.1 &44.34 &yes\\ 
SDSS J$172419.89+551058.8$ &0.365 &  8.00 &0.16 &0.016 &0.1 &0.6 &43.93 &no
\enddata 

\tablecomments{J2000 coordinates, redshifts and the [OIII]5007 luminosities (given as log($L$([OIII]5007)/$L_{\odot}$)) were taken from Paper I. Offsets (in arcmin) are between the SDSS and the RASS positions. Hardness ratios HR1 and HR2 are defined in \citet{voge99}. $L_X$ is in erg sec$^{-1}$; it is calculated assuming a power-law spectrum with $\Gamma=0$ modified by the Galactic absorption. This is a lower limit on the intrinsic X-ray luminosity, since no intrinsic absorption has been corrected for.}
\end{deluxetable}

\end{document}